\documentclass{elsart}

\usepackage[round,colon]{natbib}
\usepackage{graphicx}

\usepackage{amssymb}

\begin{document}

\begin{frontmatter}



\title{Leaking method approach to surface transport in the Mediterranean Sea from a numerical ocean model}


\author[addr1]{Judit Schneider\thanksref{email1}}
\thanks[email1]{\tt jotes@agnld.uni-potsdam.de}
\author[addr2]{Vicente Fern\'andez\thanksref{email2}}
\thanks[email2]{\tt v.fernandez@uib.es}
\author[addr2]{and Emilio Hern{\'a}ndez-Garc{\'\i}a\thanksref{email3}}
\thanks[email3]{Corresponding author: {\tt emilio@imedea.uib.es}}

\address[addr1]{University Potsdam, Institute for Nonlinear Dynamics, Potsdam, Germany}
\address[addr2]{Instituto
Mediterr\'aneo de Estudios Avanzados IMEDEA (CSIC-UIB)
\thanksref{www}, Campus de la Universitat de les Illes Balears,
E-07122 Palma de Mallorca, Spain.}
\thanks[www]{\tt http://www.imedea.uib.es/}


\begin{abstract}
We use Lagrangian diagnostics (the {\sl leaking} and the {\sl
exchange} methods) to characterize surface transport out of and
between selected regions in the Western Mediterranean. Velocity
fields are obtained from a numerical model. Residence times of
water of Atlantic origin in the Algerian basin, with a strong
seasonal dependence, are calculated. Exchange rates between these
waters and the ones occupying the northern basin are also
evaluated. At surface, northward transport is dominant, and
involves filamental features and eddy structures that can be
identified with the Algerian eddies. The impact on these results
of the presence of small scale turbulent motions is evaluated by
adding Lagrangian diffusion.

\end{abstract}


\date{26 July 2004}

\end{frontmatter}


\section{Introduction}

Ocean currents transport water in such a way that parcels
initially close become distant after some time ({\sl dispersion}).
Eventually water masses of different origins are put into contact,
allowing them to interchange their contents of heat, chemicals or
nutrients ({\sl mixing}). These two processes, dispersion and
mixing, have a profound impact on the physical, chemical and
biological dynamics of the ocean. The rapid development of
Lagrangian techniques \citep{Mariano} is enhancing our
understanding of them. For example, floater experiments, that
directly track the motion of the water parcel on which the floater
has been deployed, reveal irregular trajectories and strong
dependence of them on the precise position and timing of the
release. The influence of coherent structures on the trajectories,
such as trapping by eddies, is also observed.

One of the theoretical frameworks that is being used to understand
the above phenomena comes from the mathematical study of dynamical
systems \citep{Wig}. This approach identifies the main dynamical
structures that provide the skeleton organizing the full set of
particle trajectories. Irregular trajectories and sensitive
dependence on initial conditions are interpreted as the direct
manifestation of the chaotic nature of fluid trajectories in the
ocean. Coherent structures such as eddies and fronts are also
linked to well studied objects in the dynamical systems approach.

Although techniques for the identification of dynamical structures
in time-periodic velocity flows are well established since long
time ago \citep{Ottino,Wig}, the consideration of aperiodic flows,
as occurring in the turbulent ocean, is much more recent. A
variety of methods are being tested on more or less idealized
dynamical systems \citep{HP,DoubleGyre} and some of them applied
to realistic geophysical settings
\citep{Lacorata,Kuz,JoLe,dovidio}. A particularly simple
methodology was proposed in \citet{STN} and applied to idealized
geophysical flows \citep{ST}. The so-called {\sl leaking method}
consists in identifying the initial positions of particles that
leave a given spatial region within a long enough interval of
time. Alternatively, starting positions of trajectories which end
in a given region can also be identified. In both cases, the
geometrical structures of these sets of positions are related to
objects of dynamical relevance. In the context of chaotic
dynamical systems, they are shown to have a fractal structure,
i.e. fine details with self-similar properties at arbitrarily
small scales. The meaning of this in the ocean context is that
these sets will show a fine filamentary structure which resembles
satellite observations of sea surface temperature or ocean color
images. From a practical point of view, the fine structure is the
responsible for the different fate of trajectories starting in
close and apparently equivalent positions. The structures revealed
by the leaking method allow the identification of the {\sl
transport routes} between ocean regions, as well as the {\sl
barriers to transport}. The method was originally developed to
display the geometry of the transport structures. But it also
gives readily the relevant time scales for escape from selected
regions or residence times, an information of evident
oceanographic interest \citep{Buffoni}.

In the present paper, we apply the leaking method to a realistic
surface velocity field of the Mediterranean Sea obtained from a
primitive equation ocean model. We focus in the Western
Mediterranean, and characterize surface water exchange between a
northern and a southern sub-basins, which are areas occupied by
water masses of different characteristics.

The Paper is organized as follows: In Sect.
\ref{sec:mediterranean} we summarize the essential details of the
Mediterranean Sea oceanography needed for the rest of the Paper,
and we define the regions analyzed. Sect. \ref{sec:model}
introduces the numerical model and Sect. \ref{sec:leaking}
describes the methodology used, i.e. the {\sl leaking} and the
related {\sl exchange} methods. Residence times and Lagrangian
structures for leaking out of the southern part of the Western
Mediterranean are presented in \ref{subsec:residence}, and rates
of north-south exchange and associated structures in
\ref{subsec:NS}. In Subsection \ref{subsec:diffusion} we explore
what modifications of the above results would be induced by
unresolved small-scale turbulence, modelled here as a Lagrangian
diffusion process. The final Section contains our conclusions.

\section{Mediterranean Sea background}
\label{sec:mediterranean}

The Mediterranean is a mid-latitude semi-enclosed sea connected
with the Atlantic Ocean via the Strait of Gibraltar. It is
composed of two principal basins, the western and the eastern
basins which are connected by the strait of Sicily. The main
oceanographic characteristic of the Mediterranean Sea as a whole
is that it is an evaporative basin (evaporation exceeds
precipitation), and the deficit of water is supplied by the inflow
of Atlantic waters from the Strait of Gibraltar. This is
considered to be the main forcing mechanism of the Mediterranean
circulation. In a simplified view of the surface circulation,
Atlantic waters coming from the Strait of Gibraltar flow eastward
along the North African coast, forming the so called Algerian
current. When this current reaches the Strait of Sicily, it
bifurcates. One branch follows a cyclonic (anticlockwise) path
around the Western Mediterranean Basin, while the other branch
crosses the Strait of Sicily into the eastern basin where it also
follows a general cyclonic path. The longer the water stays in the
Mediterranean, the saltier it becomes due to mixing with adjacent
water masses and to evaporation.

The Eastern and Western Mediterranean basins themselves can be
divided in several sub-basins, which are separated by topographic
features (straits or channels) and clearly characterized by
different water masses, patterns of circulation and characteristic
forcing mechanisms (e.g heat fluxes, wind stress, rivers
discharge, etc). Indeed, each of the sub-basins present a
variability dependent on the local forcing variability and on its
own dynamics. This fact makes the Mediterranean surface
circulation to be depicted as different interacting scales,
including large scale, sub-basin and mesoscale. The interaction
between the local (sub-basin) circulation and the general
circulation, and the connections and water transports between
sub-basins, is still an open field of research of the oceanography
in the Mediterranean Sea \citep{Astraldi}.

\begin{figure}[tbp]
\includegraphics[scale=.8]{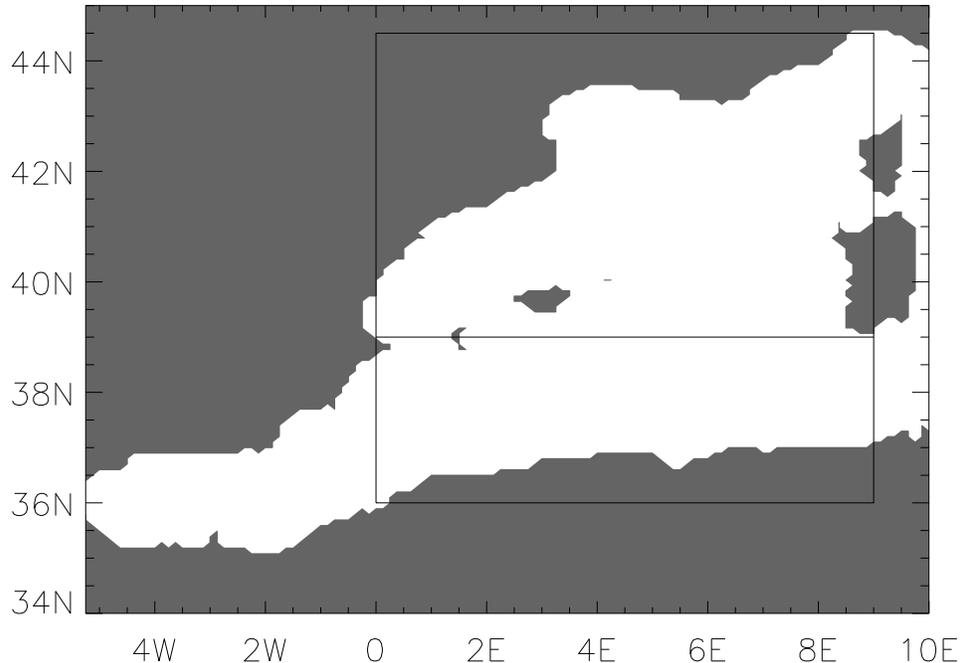}
\caption{\small The Western Mediterranean. The coastline is nearly
identical to the 16 m isobath, which is the one used by the
numerical model. The two rectangular boxes are the regions among
which transport will be addressed. The southern box contains
relatively fresh water of Atlantic origin, and the northern one
Mediterranean saltier one. } \label{fig:map}
\end{figure}

We concentrate in this Paper in the western basin, and
characterize surface transport processes involving the relatively
fresh water of Atlantic origin that fills the southern part of the
basin: its residence time and its interchanges with the saltier
water in the north. Fig. \ref{fig:map} displays the Western
Mediterranean. The two rectangular boxes are associated to the
north and south regions containing mainly the two different water
masses. The southern box extends from 36N to 39N, and the northern
one from 39N to 45.5N. The longitudinal extent of both is from 0
to 9E. The location of these geographical boundaries is arbitrary
to some extent, but we have selected the boundary between the two
regions to be at the same latitude as the Ibiza channel (the pass
between Ibiza, the most occidental island of the Balearic cluster,
and the Spanish coast) where it is known than an intense water
exchange occurs \citep{Pinot}.

\section{Mediterranean Sea numerical model}
\label{sec:model}

In this study, we have used the DieCAST (\textit{Dietrich for
Center Air Sea Technology}) numerical ocean model applied to the
Mediterranean Sea with realistic coastlines and topography.
DieCAST \citep{Dietrich} is a primitive equation, z-level, finite
difference ocean model using the hydrostatic, incompressible and
rigid lid approximations.

We summarize here some of the main properties of the model setup
(see \citet{Fernandez} for more details). The horizontal
resolution is $1/8$ degrees while the vertical direction is
discretized in $30$ levels. Horizontal eddy viscosity and
diffusivity are specified to have constant values of $A_h=K_h= 10\
m^2s^{-1}$, and are represented by a Laplacian operator. The use
of these rather small values of dissipation in DieCAST is possible
by the use of fully fourth-order-accurate numerics with special
care to reduce numerical dispersion \citep{Dietrich}. It is
noticeable that these values coincide with the value of the eddy
diffusivity estimated by \citet{Okubo} from measurement of tracer
dispersion at several scales. Okubo results can be summarized in
his empirical formula for the diffusivity (in $m^2s^{-1}$) as a
function of spatial scale $l$ (in $m$): $K_h=2.055\times 10^{-4}\
l^{1.15}$. This gives $10 m^2s^{-1}$ for our horizontal grid
resolution ($l=\Delta x \approx 12 \times 10^3 m$ is the
approximate spacing corresponding to 1/8 of degree at
Mediterranean latitudes). Vertical viscosity and diffusivity are
based on \citet{Pacanowski}, with background vertical viscosity
and diffusivity set at near-molecular values (0.01 and 0.002 $cm^2
s^{-1}$, respectively). In order to focus on the transport
mechanisms intrinsic to the internal dynamics of the
Mediterranean, we force the model with a perpetual year using a
monthly climatological wind stress and monthly climatological sea
surface temperature and salinity \citep{Dietrich2}. Tracer and
momentum time step is 15 minutes. Model outputs are stored as
instantaneous values each model day.

In this climatological configuration, the model has been able to
reproduce the correct general circulation of the basin and the
major circulation features (see \citet{Fernandez}). We used for
this study the instantaneous model Eulerian velocity data from the
second model level (representing the velocity field of a layer of
11.6 $m$ of thickness, centered at a depth of 16 $m$). Using the
second layer avoids the rather direct wind influence suffered by
the first one. The data correspond to four years ($18$ to $21$) of
a simulation taken after the model has reached an equilibrium
state for the surface circulation.

We consider just the horizontal components of the velocity.
Lagrangian motion in the horizontal is the adequate framework to
compare with usual Lagrangian measurements, in which typical
oceanographic buoys are transported by the flow while remaining at
a fixed depth. Vertical velocities in the ocean are about three or
four orders of magnitude smaller than the horizontal ones, so that
the horizontal flow is nearly divergenceless to a good
approximation in most of the considered areas (exceptions are the
places of deep water formation, which are not resolved by the
climatological forcing used here). Nevertheless, true fluid
particles will finally leave the considered layer in the vertical
direction at long times, so that the transport properties
established here should be considered (as floater experiments) as
a tool to explore the transport processes limited to the surface
layers.

\section{Lagrangian diagnostics - The leaking method}
\label{sec:leaking}

The leaking method \citep{ST,STN}, originally developed to
visualize Lagrangian structures in complex flows, focus on a
finite preselected region out of the full region accessible to the
flow. One introduces at an initial moment $t_0$ a large number of
fluid particles in this region and lets them to be transported by
the flow for a time span $\tau$, definitely longer than the
natural time scale of the flow. The particles are only followed
until they exit the selected region. Thus, the initial ($t=t_0$)
positions of these trajectories define the location of long
lifetimes inside the region. These initial positions either align
in filamental features that can be shown to identify contracting
directions in the flow \citep{ChaoticScatt}, or they delineate
coherent structures such as eddies. These structures are called
the stable manifold of the invariant set or simply the {\sl stable
manifold} \citep{ChaoticScatt}. The final ($t=t_0+\tau$) positions
of the same trajectories fall also either on coherent structures
or on filaments along which trajectories are about to escape the
preselected region. The tangents of this {\sl unstable manifold}
correspond to the local stretching directions
\citep{ChaoticScatt}. The shape and location of the obtained
structures are functions of time ($t_0$ for the stable and $t_f$
for the unstable manifold) as long as the flow is time-dependent.
The resolution of the filamental structures depends in addition on
the initial number of tracers used (from more particles finer
structures can be recovered) and the length of simulated time
$\tau$ (finding the appropriate simulation time is a crucial
point: simulating for too short times, not all particles will be
aligned along the contracting/stretching directions, whereas when
simulating for a too long time the fine filamental structures fade
out and disappear because of the decreasing number of particles
that trace them). In the cases in which the transported particles
undergo standard chaotic motion, the number of fluid particles
$N(t)$ that have not yet exit the region after a time $t \in
(t_0,t_f)$ decays exponentially $N(t) \approx N(t_0)\e^{\kappa
(t-t_0)}$ \citep{ChaoticScatt,JTZ}, from which a escape rate
$\kappa$ can be measured. Its inverse has the meaning of a {\sl
residence time} in the region. In the presence of coherent
structures, however, this exponential decay slows down into a
power-law at long times. Although we are restricting to horizontal
motion in this Paper, we mention that the leaking method can be
applied also to more complex threedimensional situations
\citep{TSPT}.

In addition to the original leaking method, we propose here a
modified version of it (that we can call the {\sl exchange}
method) in order to get additional information about exchange
rates and transport structures between two sub-basins: we want to
consider two different regions (the northern and southern boxes in
the Western Mediterranean map of Fig. \ref{fig:map}) and calculate
the transport properties between them. One of the two regions will
be filled up randomly with particles. Then, the motion of the
particles is calculated. In contrast to the original leaking
method, particles are free to leave their starting region and to
move in the whole Mediterranean. But at the moment in which they
(eventually) enter the second region, the simulation of their
motion is stopped. These particles are then counted as
\textit{exchanged} ones. Again, the local tangents to the initial
and final distributions of non-exchanged particles identify
contracting and stretching directions, and coherent structures
(and the same applies to the distribution of exchanged ones) . In
addition, the initial positions of the exchanged particles
identify the {\sl exchange set}, i.e., the initial locations that
contribute to transport to the second area within the considered
time span. Time scales for transport among the regions can be
conveniently defined (see Sect. \ref{subsec:NS}).

Motion of particles is calculated by using the Eulerian velocity
fields, previously stored from the numerical simulation. Linear
interpolation in time is used to downsize the $1$ day interval
between stored data to the $0.5$ hour time-step of the 4-order
Runge-Kutta used in the trajectory calculation. In space, we use
bilinear interpolation to compute velocities in between the model
gridpoints.

\section{Results}
\label{sec:results}

\subsection{Exit process and residence times in the southern Western Mediterranean}
\label{subsec:residence}

We apply the leaking method to the southern box depicted in Fig.
\ref{fig:map} (comprised between 0 and 9E and between 36N and
39N). This box is traversed by the Algerian current, flowing
eastwards, and contains mainly low-salinity water masses of
Atlantic origin (S $<$ 37.5 psu). We estimate the residence times
inside the box, and the geometrical paths of escape.

\begin{figure}[tbp]
\includegraphics[scale=0.57]{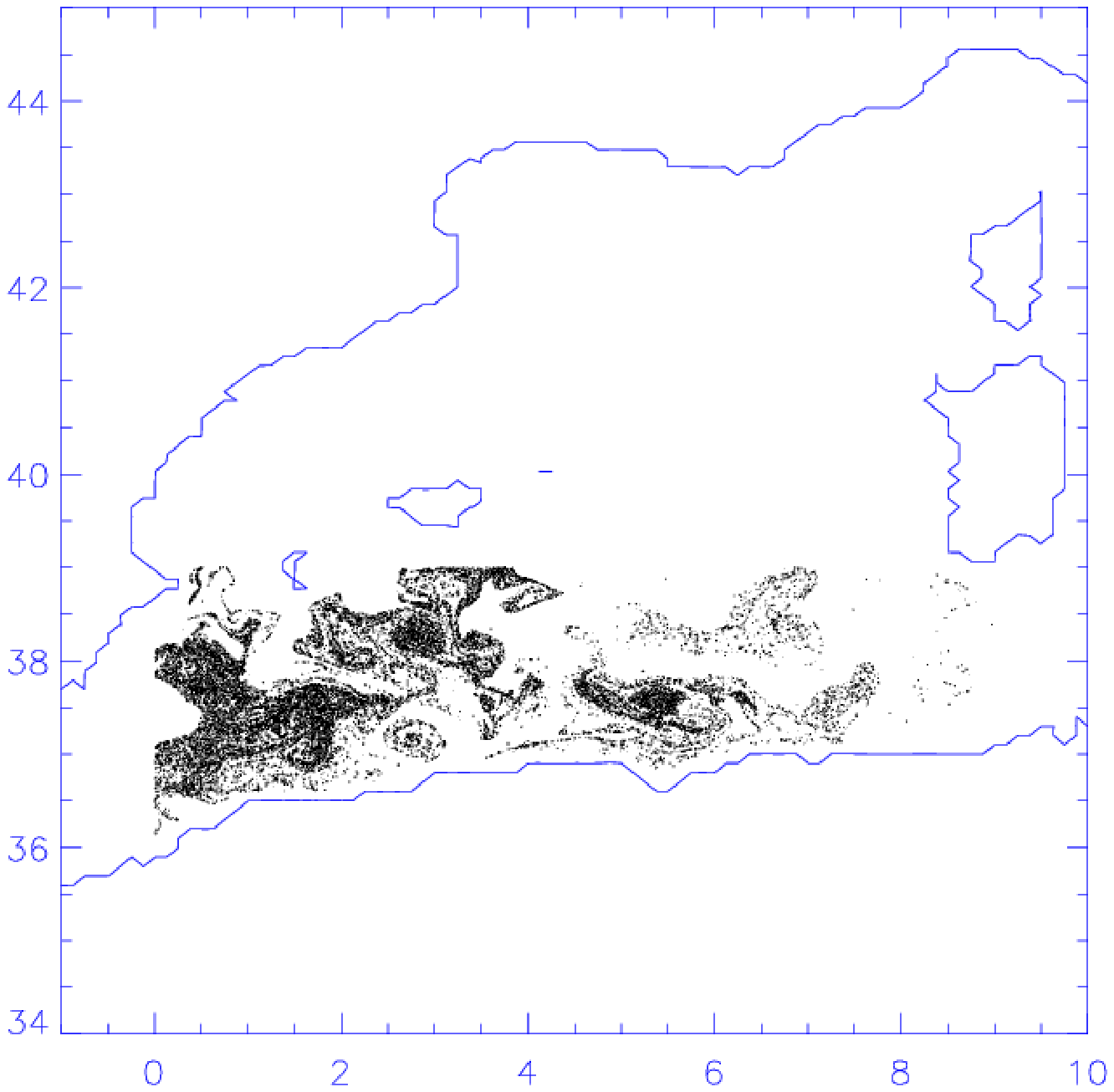}
\includegraphics[scale=0.57]{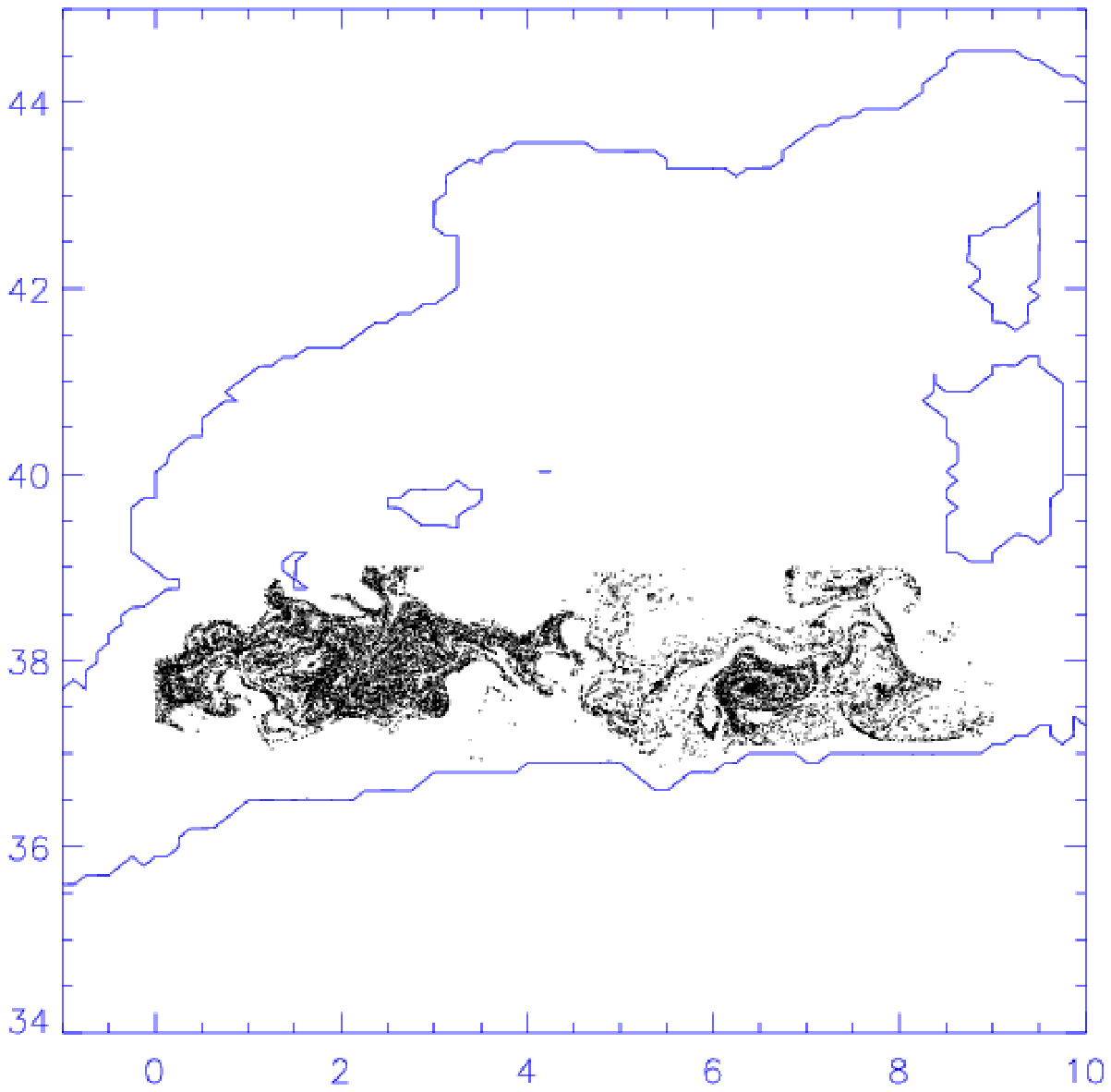}
\caption{\small Left: The initial positions of particles which did
not leave the southern box during an interval of time
$(t_0,t_0+\tau)$, with the starting time $t_0$ at the beginning of
Winter of the first simulation year and $\tau=90$ days. They trace
the stable manifold of the invariant set, and local tangents
identify contracting directions. Right: Final positions of the
above particles, tracing out the unstable manifold and the
stretching directions.} \label{fig:usm_box}
\end{figure}

We put at random initial positions a large number of particles in the box. We
then follow numerically during $\tau=90$ days the trajectories of the
$N(t_0)=122352$
particles that were not started in land. Particles are marked as
escaped when they first leave the box, and their trajectories are
no longer followed. The final positions of particles which remain
in the box until the end of the simulation depict the unstable
manifold and stretching directions, whereas their initial
positions are associated to the stable manifold and contracting
directions. Fig. \ref{fig:usm_box} shows an example of such
structures, obtained by starting the Lagrangian integrations at a
time $t_0$ corresponding to the first day of Winter
(in this paper, Winter will denote the months of January, February
and March, i. e. the first 90 days of the year, Spring the next
three months, Summer the following ones, until Autumn which will
comprise October, November and December).

We see complex geometrical structures with filamental features and
also the presence of eddies. The initial location of non-escaped
particles is far from compact and is interleaved with the initial
location of the escaped ones, so that small displacements in this
initial position will change the fate of the particle. We see that
escape is more frequent when starting in the eastern part of the
box, whereas particles started in the west have more chances to
become trapped in structures suggestive of Algerian eddies.

\begin{figure}[tbp]
\includegraphics[scale=0.57]{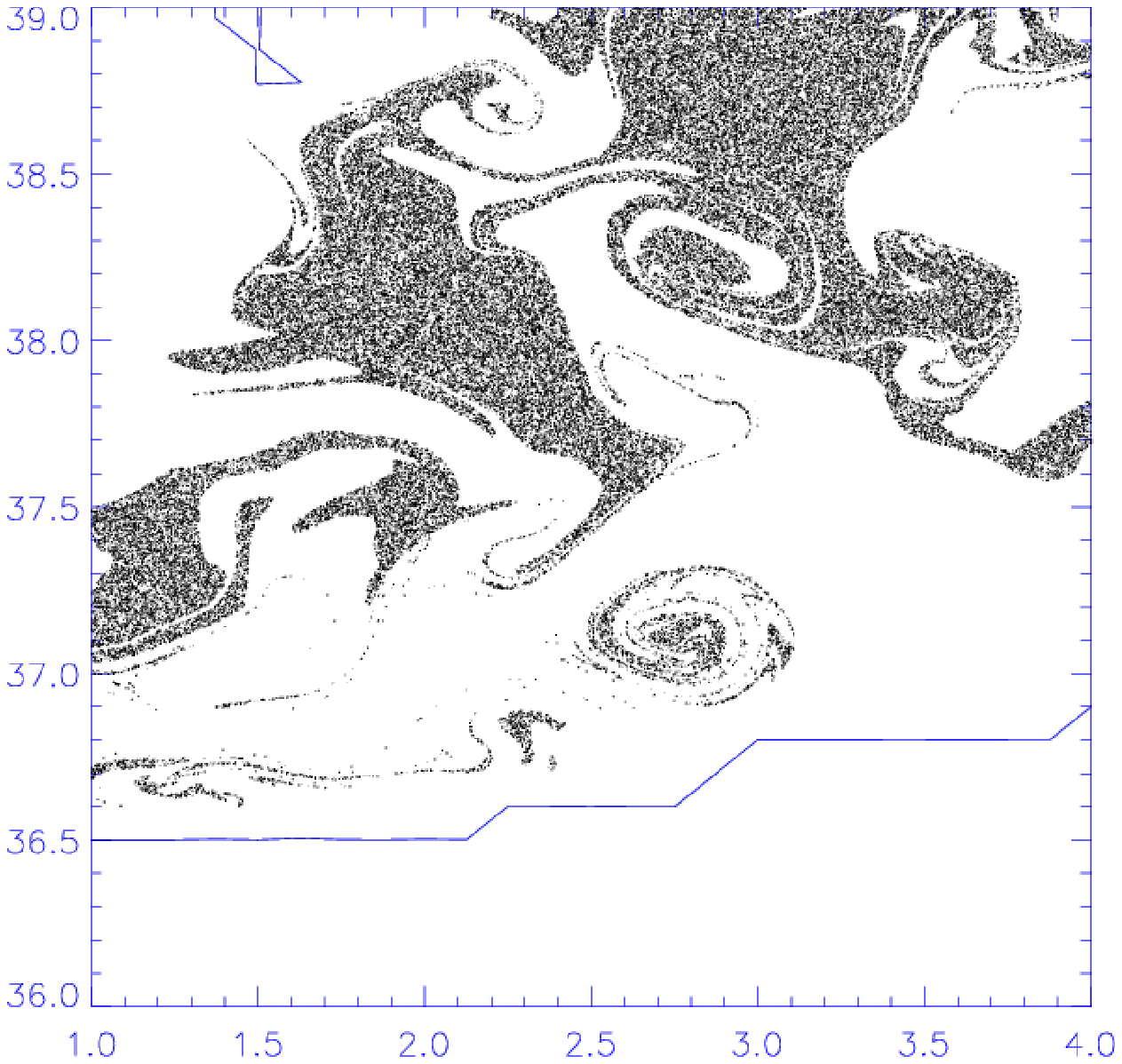}
\includegraphics[scale=0.57]{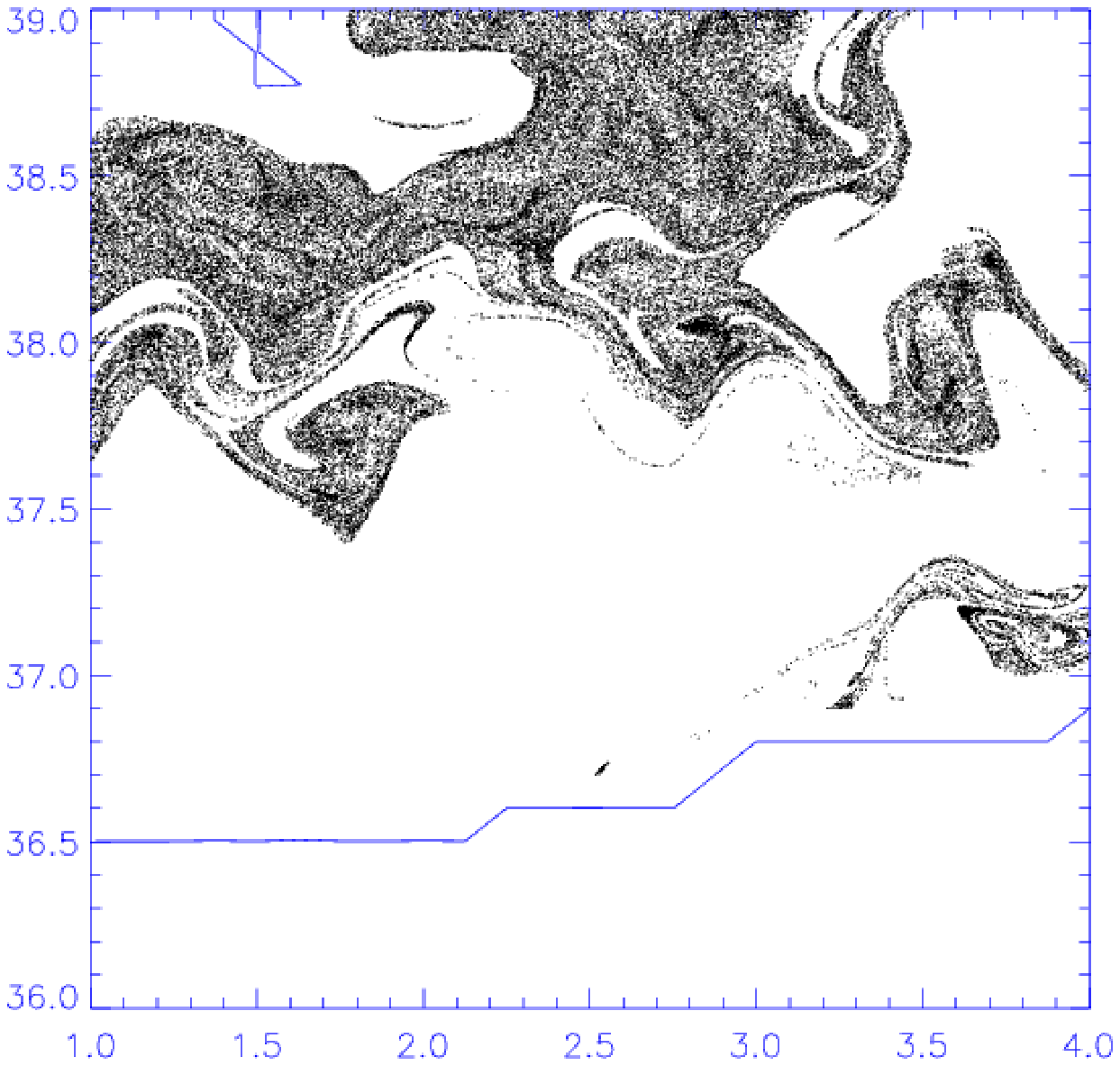}
\caption{\small Idem as Fig. \ref{fig:usm_box}, but using a
smaller box in the leaking approach, and an integration time of
$\tau=30$ days. Qualitatively, it can be considered as a blow-up
of Fig. \ref{fig:usm_box}.} \label{fig:detailed_usm_box}
\end{figure}

The fine structure of the geometry of the escaping and
non-escaping sets is better appreciated in Fig.
\ref{fig:detailed_usm_box}. It can be though qualitatively as a
blow up of Fig. \ref{fig:usm_box}, although, for clarity, it has
been produced in a different way: a number of particles has been
integrated for $\tau=30$ days, but considering the leaking from
the smaller region shown in Fig. \ref{fig:detailed_usm_box}.
Although different in detail, the main features should be similar
to the ones in Fig. \ref{fig:usm_box}. Filaments, eddies, and the
entanglement between escaping and non-escaping positions, are
particularly clear here.

\begin{figure}[tbp]
\includegraphics[angle=-90,scale=0.45]{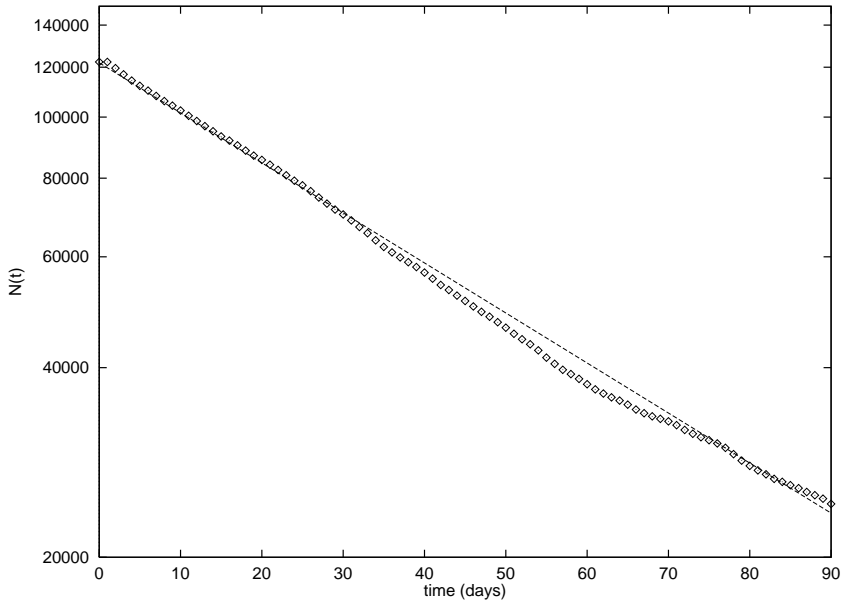}
\caption{\small
The number of particles that did not leave the southern box in
Fig. \ref{fig:map} after being deployed at the beginning of Winter
in the first simulation year (symbols). The escape rate is
obtained by an exponential fitting (dashed line). }
\label{fig:escape_box}
\end{figure}

The escape rate from the original [0,9E]$\times$[36N,39N] box can
be calculated from the decay of $N(t)$, the number of non-escaped
particles after a time $t$. For the same example of launching the
particles in Winter, as before, Fig. \ref{fig:escape_box} displays
a clear exponential decay during the integration time, so that a
escape or leaking rate ($1.93\cdot 10^{-2} \ days^{-1}$ for the
Winter example of the figures)
can be obtained from an exponential fitting. The inverse of the rate can be
interpreted as a residence time inside the region (51.8 days for the example
before). Table \ref{table:leaking} shows the seasonal dependence of the rates,
with the faster escape in Autumn and the slower in Summer. The yearly averaged
residence time in the southern basin is 59.4 days. This residence time is
comparable to estimations based on lagrangian data, which gives a few months on
average for a particle to escape from the Algerian basin \citep{millot91,salas}.

Seasonal cycle is the main component of variability in the leaking
rates (the model is forced by a seasonal atmospheric cycle), but
note also the existence of a signal of interannual variability,
which reflects non-forced interannual changes in ocean surface
mesoscale structures due to the non-linear character of the ocean
dynamics.

\begin{table}[btp]
\begin{tabular*}{0.67\textwidth}{@{\extracolsep{\fill}}cccr}
Season & leaking rates (days$^{-1})$ & (std (days$^{-1}$)) \\
\hline \\
Winter  & $1.93 \cdot 10^{-2}$ & $(1.69 \cdot 10^{-3})$ \\
Spring  & $1.38 \cdot 10^{-2}$ & $(1.55 \cdot 10^{-3})$ \\
Summer  & $1.39 \cdot 10^{-2}$ & $(2.35 \cdot 10^{-3})$ \\
Autumn  & $2.02 \cdot 10^{-2}$ & $(1.54 \cdot 10^{-3})$ \\
\\
Yearly Average & $1.68 \cdot 10^{-2}$ & $(3.40 \cdot 10^{-3})$
\\
\end{tabular*}
\caption{\small Leaking rates out of the southern box for each
season, averaged over $4$ years, and the yearly average, with the corresponding
standard deviation (std) characterizing the dispersion among the four years.
Rates are in $1/day$.}
\label{table:leaking}
\end{table}

\subsection{North-South exchange}
\label{subsec:NS}

We focus here on transport between the northern and southern regions of the
western Mediterranean, by using the modified leaking method, or exchange method.

There is observational evidence of an intense interchange of
surface waters between these two regions, in which northern
saltier waters are transported to the south to close the cyclonic
western Mediterranean circulation, but also southern fresher
waters are found in the north \citep{Pinot,pascual}. The routes of
transport for this exchange, specially for the northward
transport, are not still very well defined.

A particle started in the south is considered to be exchanged to
the north when it first hits the northern box, and viceversa.
Exchange during $\tau=90$ days is nearly equivalent to a simple
crossing of the 39N parallel. This is so because o f the location
of the boxes and of the coast, and because particles leaving the
southern box towards the east will need a time longer than
three-four months to reach the northern box moving along the
Italian coast. Therefore, with the given geometry of the
surroundings of the boxes, the exchange is comparable to a simple
crossing, and the exchange rate equivalent to a crossing rate.
Nevertheless, in the computations we use the criteria presented
when explaining the exchange method.

\begin{figure}[tbp]
\includegraphics[angle=-90,scale=0.45]{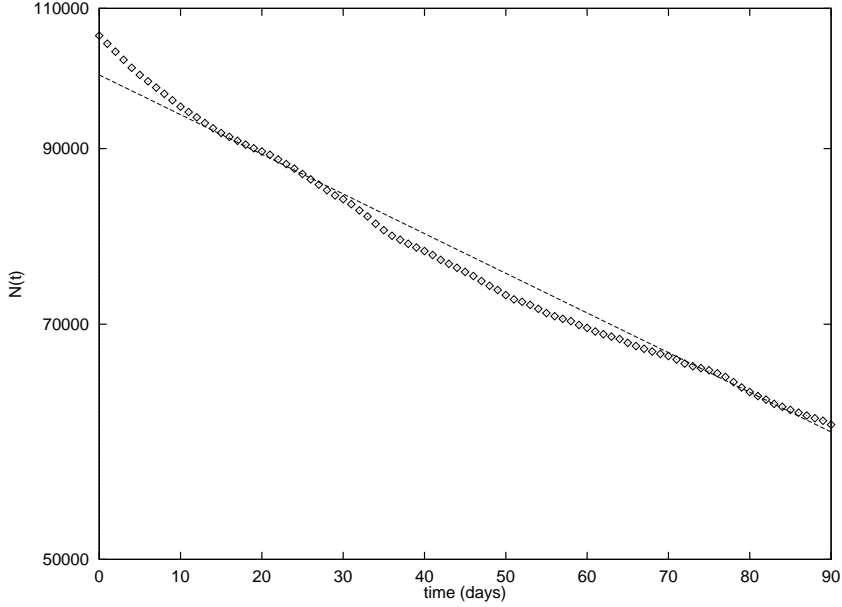}
\caption{\small The number of non-exchanged particles $N(t)$, i.e.
the ones that did not exchange to the northern box a time $t$
after starting from the southern one in Winter of the first
simulation year (symbols). The dashed line is an exponential
function decaying with the average rate $\overline{\kappa}$
defined in the text. } \label{fig:exchange}
\end{figure}

As before, we initially fill-up the southern (northern) box with a
large number of particles. $N(t_0)=105791$ ($N(t_0)=93675$) of
them did not start in land and were followed until exchange to the
other box occurred. Figure \ref{fig:exchange} shows the decay in
the number of non-exchanged particles starting from the southern
box in Winter of the first simulation year. It is clear that the
decay now is not a simple exponential. The reason is that, as time
advances, the particles are in average further and further away
from the target region, so that the rate decreases in time.
However, we can define an {\sl average} exchange rate as
$\overline{\kappa}=\tau^{-1}\ln\left(N(t_0)/N(t_f)\right)$, which
characterizes the effective rate of exchange between the two
regions during the considered integration time. For the example of
Fig. \ref{fig:exchange} we find $\overline\kappa=2.91 \cdot
10^{-3}$ days$^{-1}$. Note that, given that the inverse of this
value (343.6 days) is much longer than the integration time
($\tau=90$) we can not take it as a reliable average residence
time. The value of $\overline\kappa$ however, when multiplied by
the volume of water associated to the initial box, would give an
estimation of the effective volume transport rate towards the
second box during the integration time.

Table \ref{table:NS} summarizes the seasonal ($\tau=90$) crossing
rates $\overline\kappa$ among the two boxes, averaged over 4
years, with the corresponding standard deviations giving an
indication on the interannual variability in the crossing rates.

\begin{table}[btp]
\begin{tabular*}{0.85\textwidth}{@{\extracolsep{\fill}}ccccc}
& N to S & (std) & S to N  & (std)\\
\hline \\
Winter  & $3.06 \cdot10^{-3}$ & $(2.75 \cdot 10^{-4})$ & $6.59 \cdot 10^{-3}$ &
$(6.43 \cdot 10^{-4})$\\
Spring & $2.38 \cdot 10^{-3}$ & $(1.39 \cdot 10^{-3})$ & $5.89 \cdot 10^{-3}$ &
$(1.24 \cdot 10^{-3})$\\
Summer & $2.91 \cdot 10^{-3}$ & $(6.91 \cdot 10^{-4})$ & $6.11 \cdot 10^{-3}$ &
$(1.31 \cdot 10^{-3})$\\
Autumn & $3.03 \cdot 10^{-3}$ & $(3.18 \cdot 10^{-4})$ & $3.83 \cdot 10^{-3}$ &
$(5.96 \cdot 10^{-4})$\\
\\
Yearly Average &$2.84 \cdot 10^{-3}$  & $(3.16 \cdot 10^{-4})$  & $5.6 \cdot 10^
{-3}$ & $(1.2 \cdot 10^{-3})$ \\
\\
\end{tabular*}
\caption{\small Seasonal exchange rates
 between the two boxes (in $day^{-1}$) averaged over the 4 years, and the
 associated standard deviation (std) characterizing variability among the 4 year
s.}
 \label{table:NS}
\end{table}

We see that there is a seasonal variability of the crossing rates,
being Winter the season in which more exchange occurs in both
directions, and also the one recording the maximum imbalance
between them. This imbalance is strong during all the seasons
except Autumn, so that there is a net annual rate from south to
north, implying that surface Atlantic waters from the Algerian
basin are, on average, crossing the 39N line and mixing with
saltier waters in the north. This northward surface transport must
be compensated by a net southward transport of saltier (denser)
water in deeper layers.

\begin{figure}[tbp]
\includegraphics[scale=0.57]{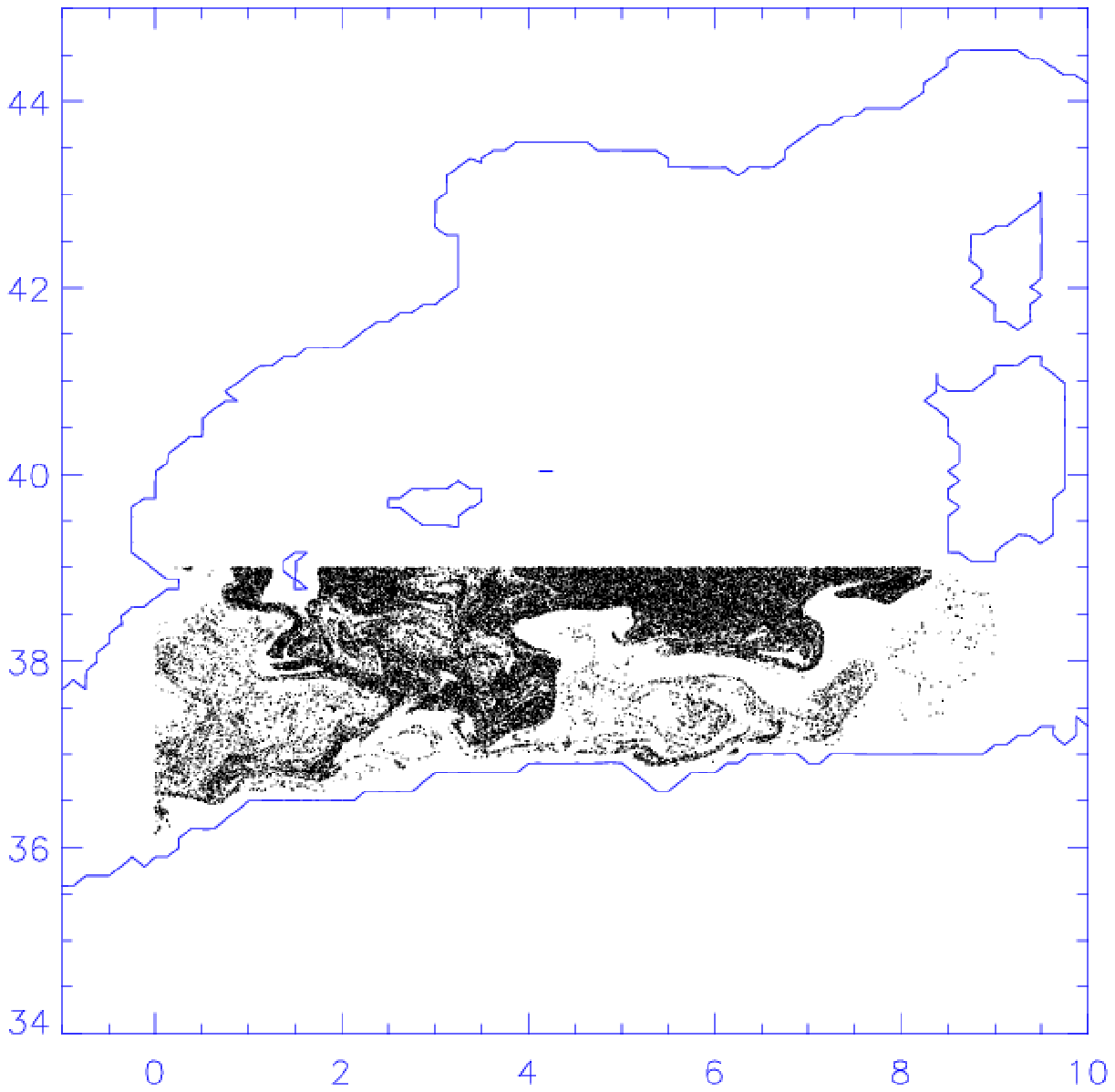}
\includegraphics[scale=0.57]{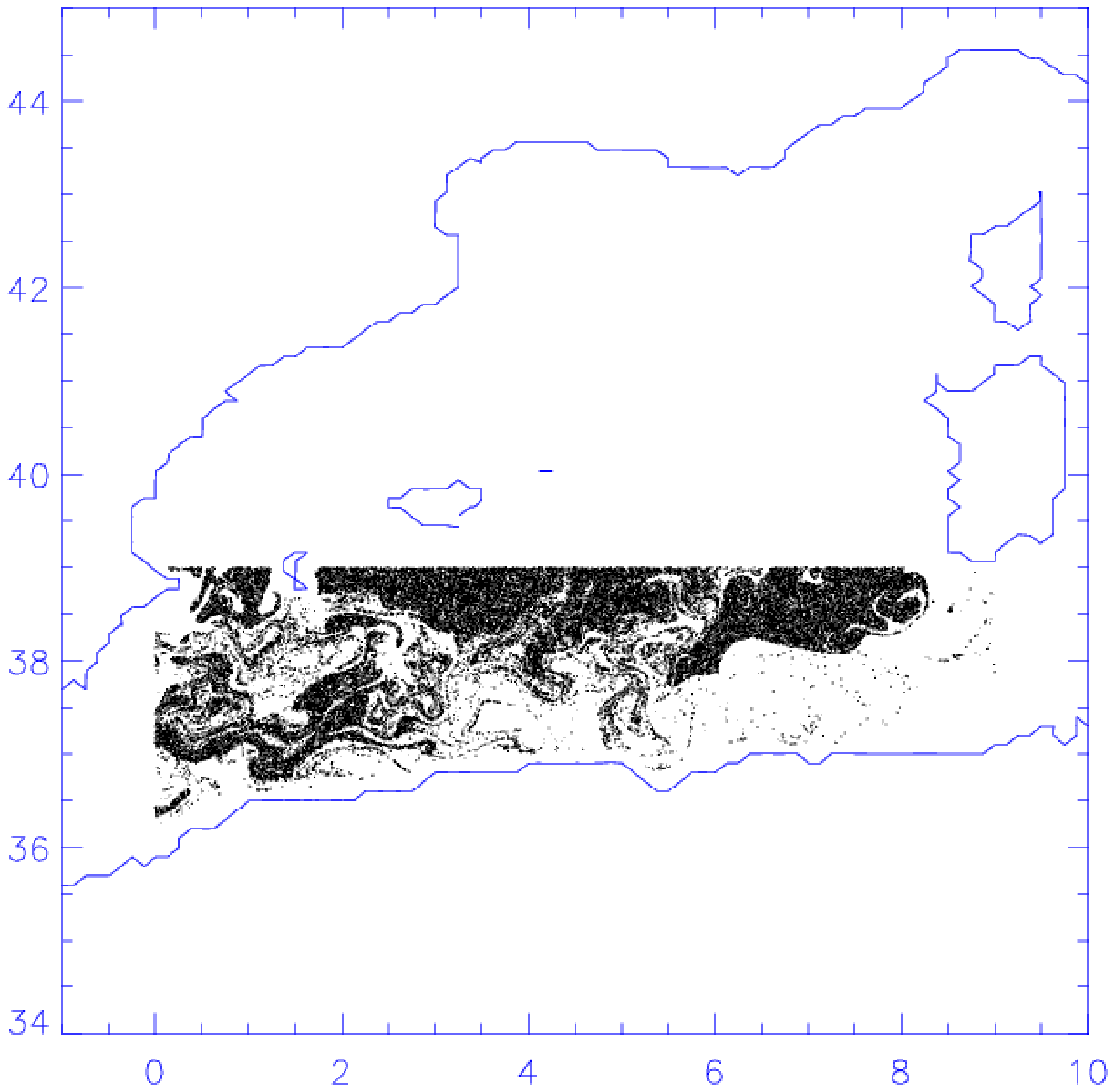}
\includegraphics[scale=0.57]{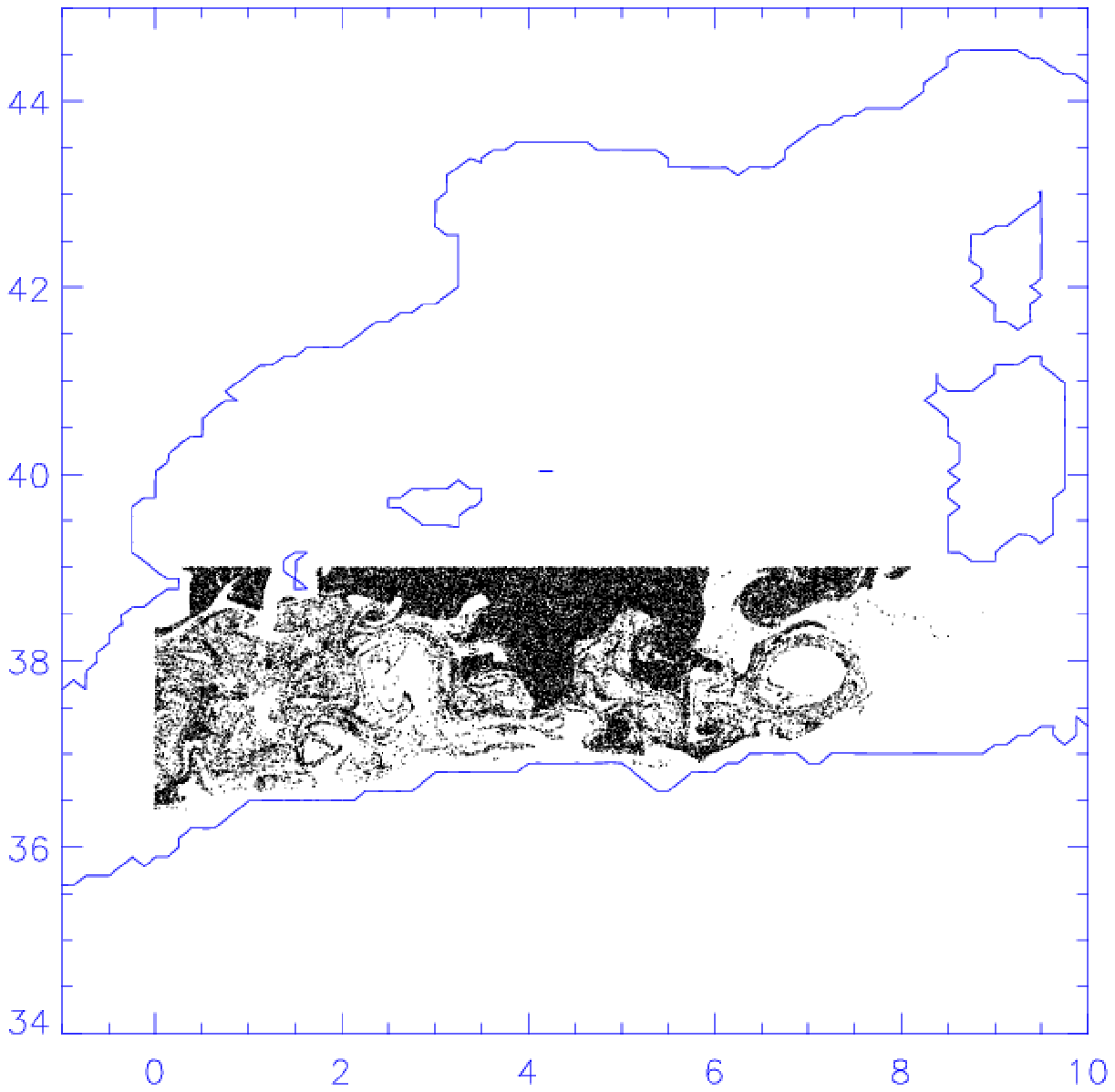}
\includegraphics[scale=0.57]{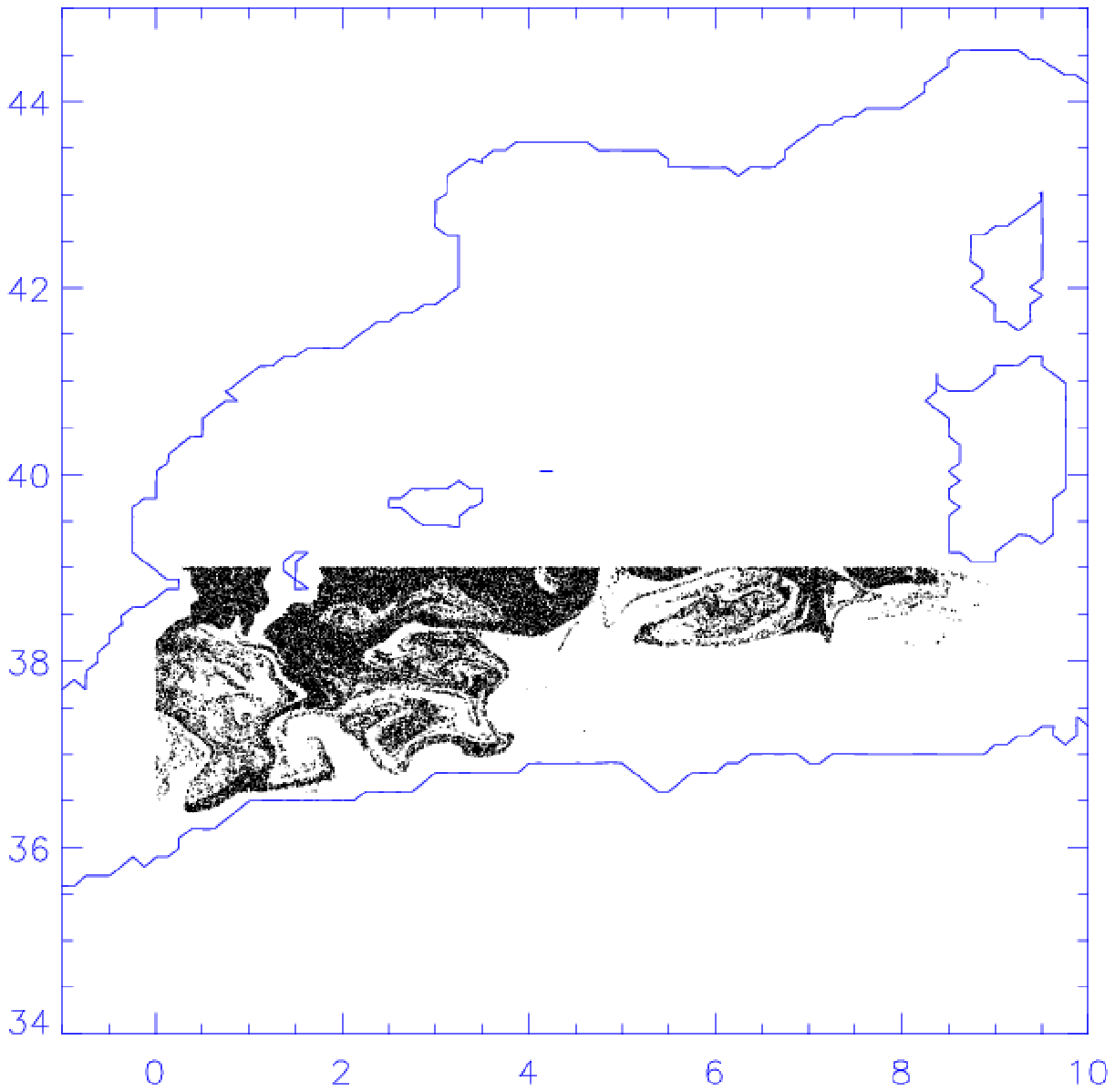}
\caption{\small Exchange sets towards the northern box starting in
the south at different seasons (first simulation year): Upper
left: Winter, upper right: Spring, lower left: Summer, lower
right: Autumn.} \label{fig:outS}
\end{figure}

To get more insight into the transport routes, we investigate the
initial and final positions of the particles released in each of
the boxes. We plot in Fig. \ref{fig:outS} the {\sl exchange set}
from south to north, i.e. the set of southern initial conditions
which bring water to the northern box during the interval of time
$\tau$. The shape of the exchange set changes each season,
according to the seasonal dependence of the exchanging or crossing
rates. The exchange set is smallest in Autumn in correspondence to
the smaller crossing rate in Table \ref{table:NS}. The exchange
set is denser close to the boundary between the regions,
indicating an easier crossing towards the north for the particles
starting there, but it extends deep into the interior of the
southern box, forming filamental structures interleaved with the
non-exchanged set. Eddy structures, specially their outer parts,
are clearly involved in the northern transport.

\begin{figure}[tbp]
\includegraphics[scale=0.8]{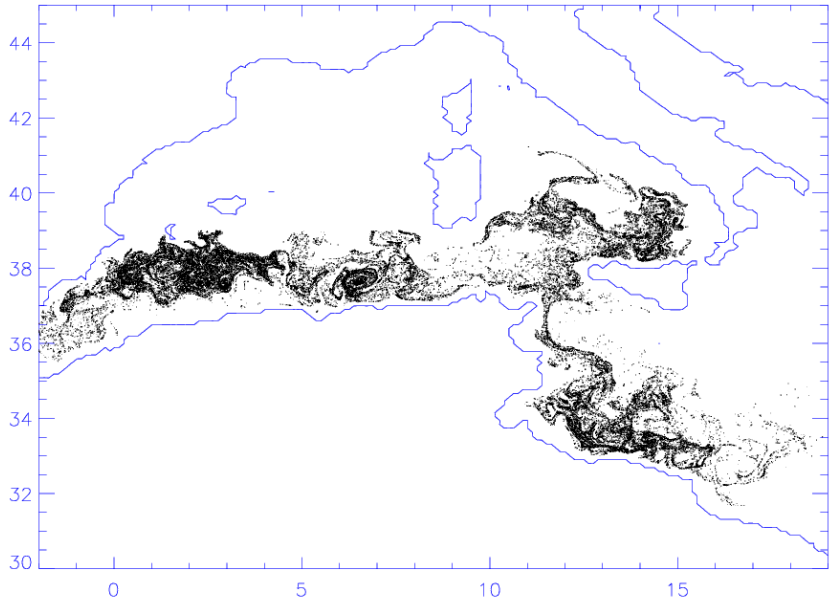}
\caption{\small
Final positions (after $\tau=90$ days) of particles launched in Winter (first
simulation year) in the southern box which did not exchange to the North.}
\label{fig:usmS}
\end{figure}

The eastern part of the Algerian region tends to be excluded from
the exchange set, specially in Autumn. Particles started there
will be transported eastward by the Algerian current. This can be
seen when plotting the final positions of the particles which did
not exchange into the northern box. As the previous pictures, Fig.
\ref{fig:usmS} shows these positions for particles started in the
southern box in Winter. Non-exchanged particles followed the
Algerian current along the African coast towards the East. Shortly
before reaching the Strait of Sicily, the current splits into two
parts: one going into the Tyrrhenian Sea (reaching the northern
box along this route would take more than the 90 days used in this
computation), and the other one following the coastal current
along the African coast into the eastern basin of the
Mediterranean. However, some particles are still in their initial
southern box. They seem to be trapped in several vortexes close to
the African coast.

\begin{figure}[tbp]
\includegraphics[scale=0.57]{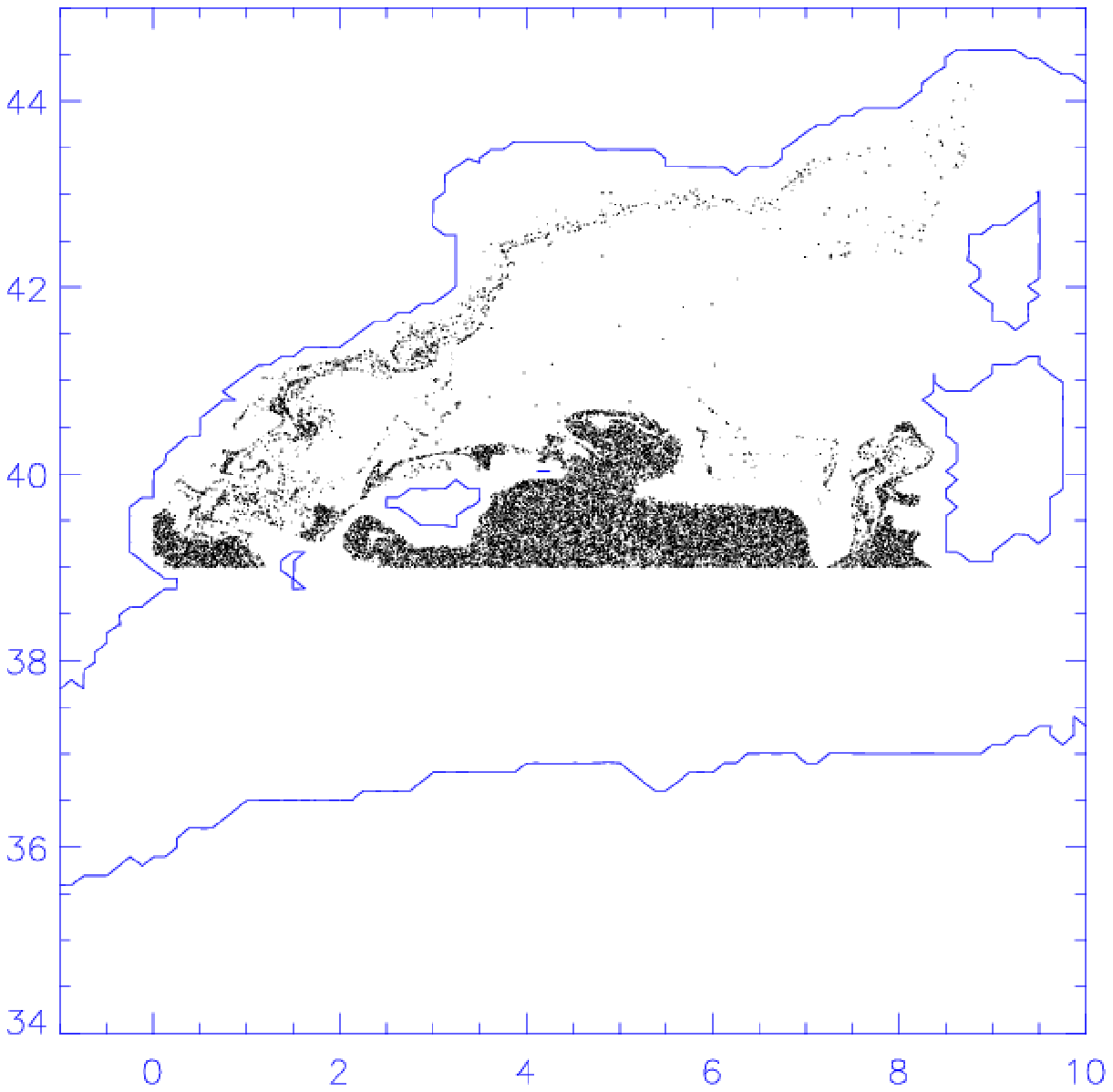}
\includegraphics[scale=0.57]{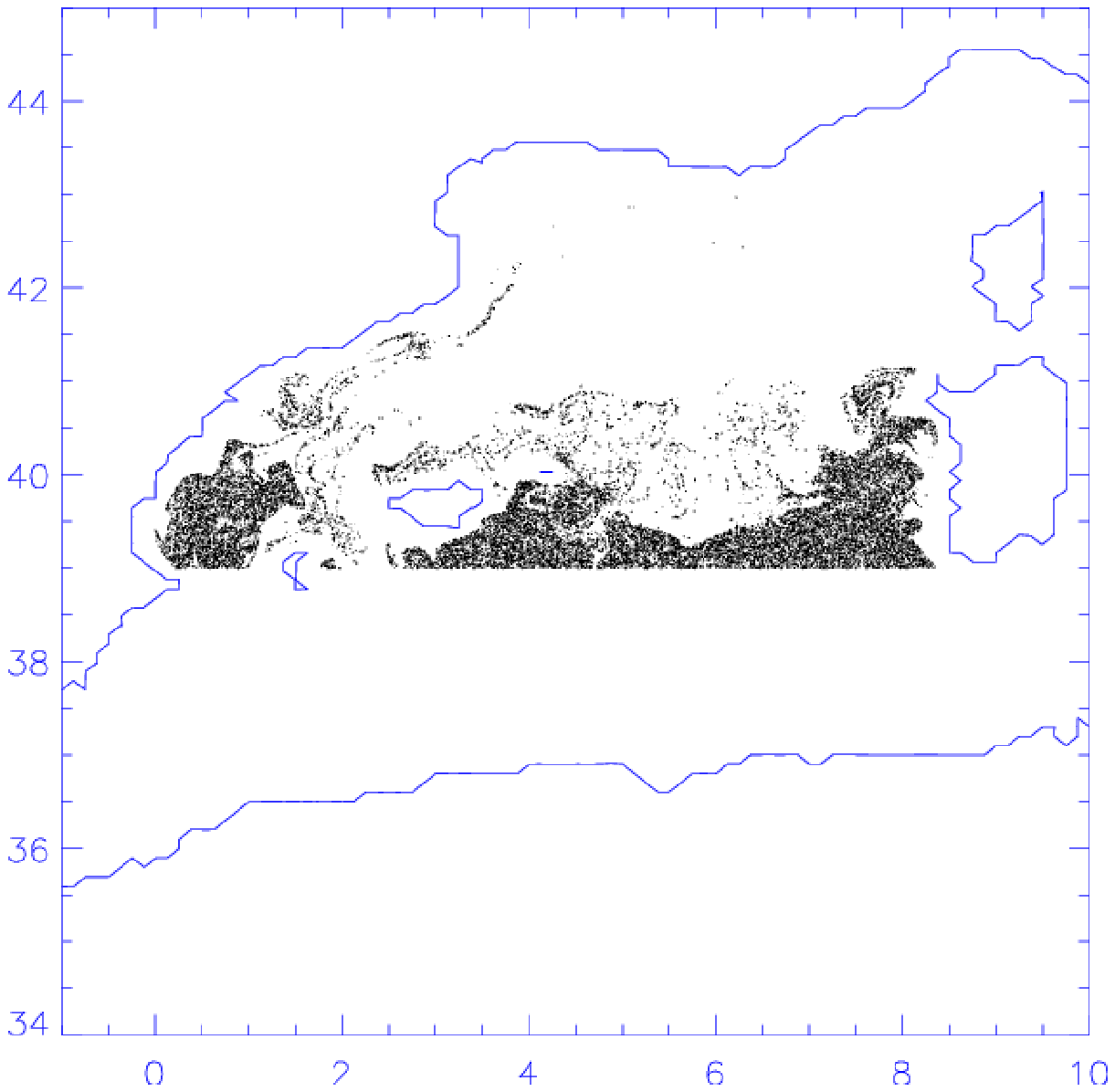}
\includegraphics[scale=0.57]{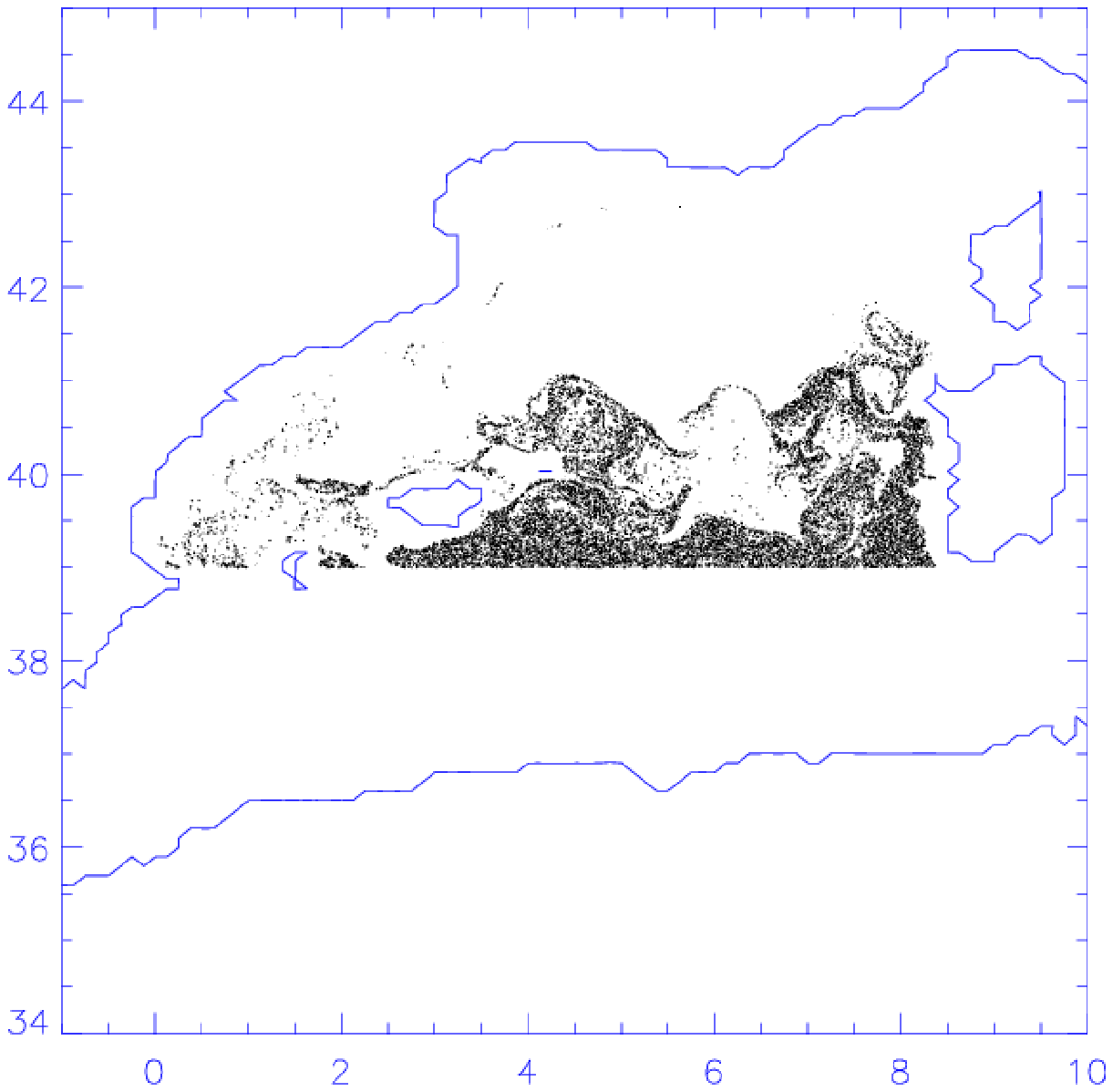}
\includegraphics[scale=0.57]{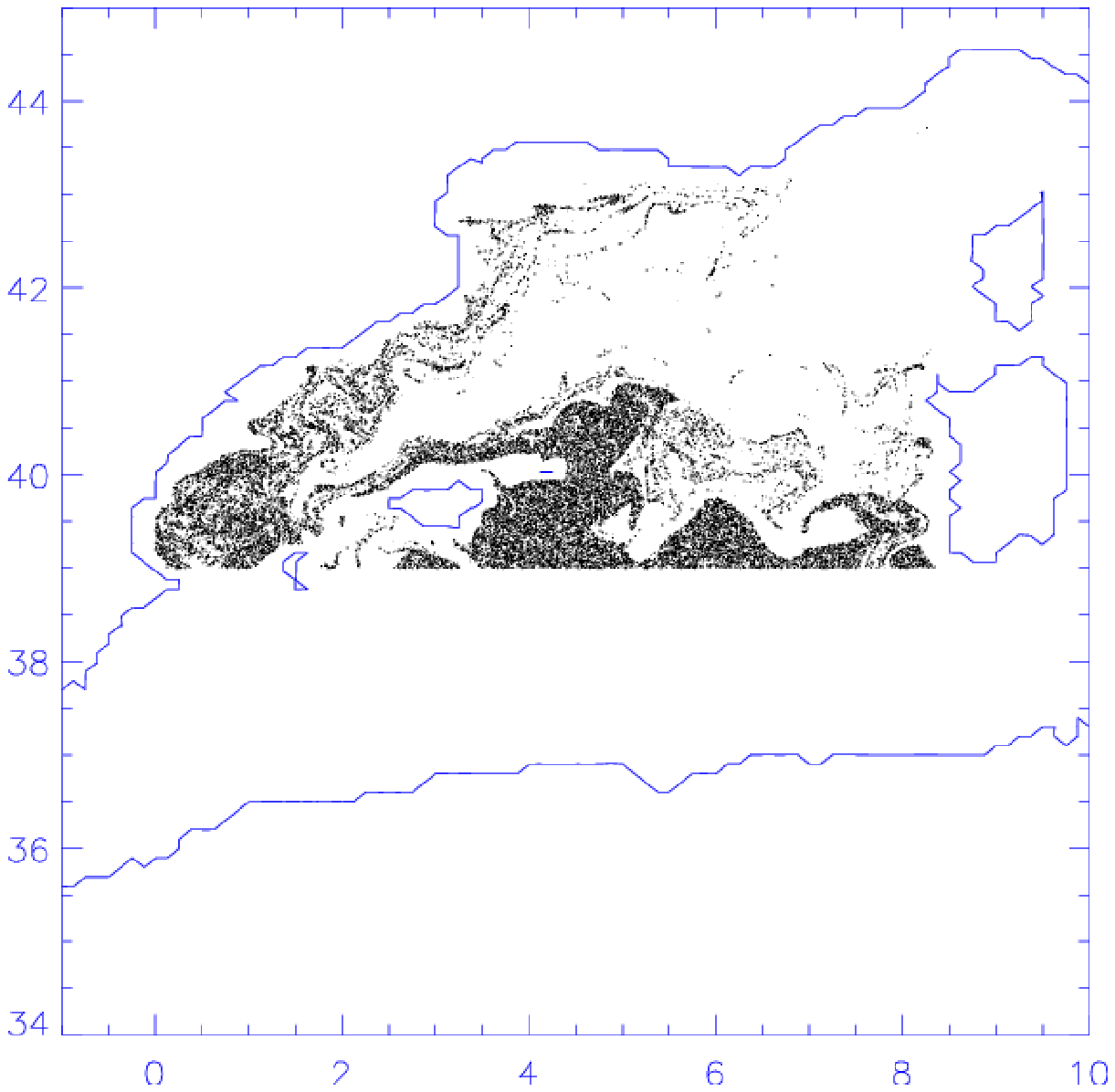}
\caption{\small Exchange sets towards the southern box starting in
the north at different seasons (first simulation year): Upper
left: Winter, upper right: Spring, lower left: Summer, lower
right: Autumn. } \label{fig:outN}
\end{figure}

Fig. \ref{fig:outN} shows another sequence of the exchange sets,
but this time towards the southern box for particles launched in
the north. In comparison with the exchange sets in Fig.
\ref{fig:outS}, here they are more restricted to the neighborhood
of the crossing boundary, except for the presence of a coastal
current along the Spanish coast (Catalan current), more visible in
Autumn and less intense in the Summer. In accordance with the
smaller crossing set, the crossing rates shown in Table
\ref{table:NS} are also smaller. At all seasons, the exchange set
excludes a region starting north of the Mallorca channel (the
strait between Ibiza and Mallorca islands) and extending
northeast. This identifies the location of a strong current
transporting particles in the northeast direction.

\begin{figure}[tbp]
\includegraphics[scale=0.8]{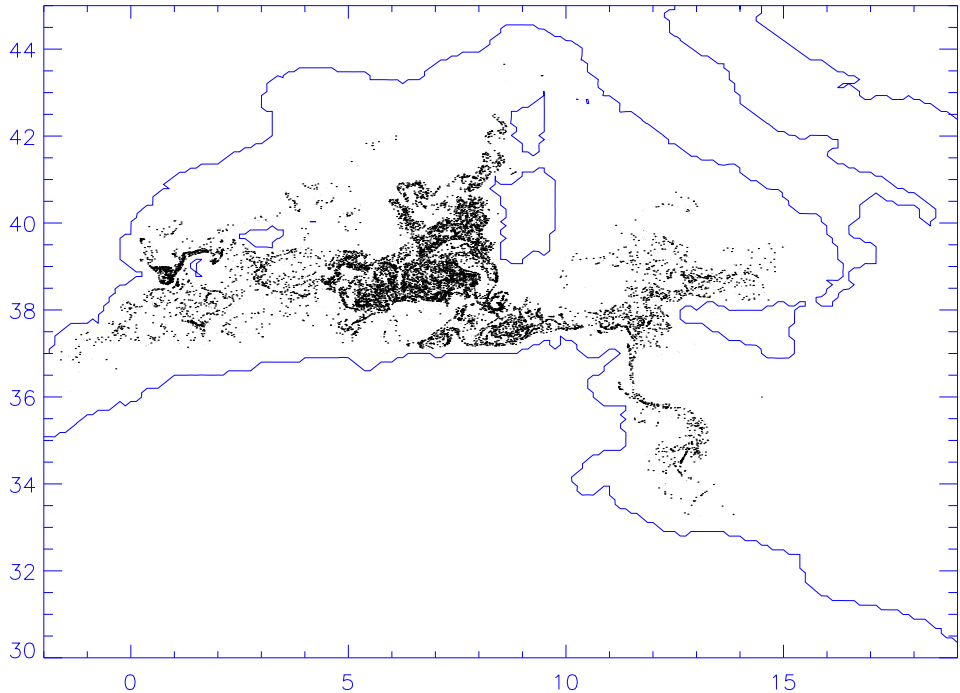}
\caption{\small
Final positions (after $\tau=90$ days) of particles launched in Winter (first
simulation year) in the northern box which did exchange to the South.}
\label{fig:umN}
\end{figure}

To get additional indications of the mixing between different
water masses, we plot in  Fig. \ref{fig:umN} final positions of
particles which exchanged from the northern towards the southern
box during the 90 days integration, starting in Winter. Many
particles are tracing out the same structures as particles
launched in the south: both are showing the routes of transport of
the main circulation in the basin, partly going to the eastern
basin of the Mediterranean. Thus mixing of water from both origins
is occurring along these routes. There are however differences.
For example there is some accumulation of particles southwest of
Sardinia island, indicating that many of the exchanged particles
remain recirculating in that area. In addition, eddy structures
are less evident here, in particular for the further reaching
particles, than in Fig. \ref{fig:outS}. From this and other
comparisons between Figs. \ref{fig:outS}, \ref{fig:usmS},
\ref{fig:outN} and \ref{fig:umN}, we can conclude that transport
mediated by vortices is more important for the south-north
direction than  for the reverse, corresponding to the abundance
and origin of these structures close to the Algerian coast
\citep{millot,salas}.

\subsection{The impact of additional diffusive processes}
\label{subsec:diffusion}

Our ocean model velocity fields, as any others obtained from
models or data of finite resolution, lack structures below the
horizontal grid size, which is 1/8 of degree (10-12 km) in our
case. Computation of Lagrangian trajectories uses spatial
interpolation to follow the particles and thus Lagrangian
geometrical features can be obtained that are smaller than this
limit (see for example Fig. \ref{fig:detailed_usm_box}). Thus
there is the question on how realistic are these small structures
and how the different rates calculated here will change when
including smaller scales in the numerical model. A convenient way
to include unresolved scales in Lagrangian computations is to add
to the velocity field experienced by the Lagrangian particle a
fluctuating term representing small-scale turbulence
\citep{Griffa96,Mariano}: $\dot {\bf x}(t)= {\bf v}({\bf x}(t),t)
+ \sqrt{2K}\Gamma(t)$. This gives additional diffusion to particle
trajectories. Usually $\Gamma(t)$ is taken to be a Gaussian Markov
process with a memory time that for Mediterranean modelling is of
the order of some days \citep{Buffoni,Falco2000}. This kind of
Markov process leads to a situation that is somehow in between the
purely deterministic calculation using the model velocity field
alone, and the larger stochasticity that is obtained when
$\Gamma(t)$ is a memoryless white noise. Here, to explore the
impact of irregular unresolved motions in the opposite extreme to
the deterministic situation considered in the previous sections,
we use for $\Gamma(t)$ a Gaussian white noise of zero mean and
correlations $\left<\Gamma(t) \Gamma(t')\right>=\delta(t-t')$. We
take for its strength $K$ the same value of the horizontal
diffusivity $K_h$ used in the model simulations $K_h=10\
m^2s^{-1}$, which in turn was the same value, as discussed in
Sect. \ref{sec:model}, estimated by \citet{Okubo}, as the
effective diffusivity at the scale of the model grid. After
integrations sufficiently longer than the memory time, no
significant differences are expected between using this
uncorrelated motion or the more realistic Markov process
\citep{Griffa96,Buffoni,Falco2000,Mariano} for the Lagrangian
diffusion.

Table \ref{table:NS_diffo} shows the recalculated exchange rates,
to be compared with table \ref{table:NS}. We see that the impact
of adding Lagrangian diffusion is not strong. In most of the cases
the effect is the intuitively expected slight acceleration of the
rates, although there is a case in which diffusion slows down the
exchange process. This will occur when diffusion kicks-off
particles from the fine structures in the main transport routes.

\begin{table}[btp]
\begin{tabular*}{0.85\textwidth}%
     {@{\extracolsep{\fill}}ccccc}
& N to S & (std) & S to N & (std) \\
\hline\\
Winter  & $3.11 \cdot10^{-3}$ & $(2.74 \cdot10^{-4})$ & $6.82 \cdot 10^{-3}$ & $
(6.59 \cdot10^{-4})$ \\
Spring & $2.45 \cdot 10^{-3}$ & $(2.61 \cdot10^{-4})$ & $6.19 \cdot 10^{-3}$ & $
(1.23 \cdot10^{-3})$\\
Summer & $3.04 \cdot 10^{-3}$ & $(6.81 \cdot10^{-4})$ & $6.38 \cdot 10^{-3}$ & $
(1.4 \cdot10^{-3})$\\
Autumn & $2.77 \cdot 10^{-3}$ & $(7.22 \cdot10^{-4})$ & $4.07 \cdot 10^{-3}$ & $
(6.53 \cdot10^{-4})$\\
\\
Yearly Average & $2.84 \cdot 10^{-3}$ & $(2.99 \cdot 10^{-4})$ & $5.87 \cdot 10^
{-3}$ & $(1.2 \cdot 10^{-3})$ \\
\\
  \end{tabular*}
\caption{\small Exchange rates (in $day^{-1}$, averaged over the 4
simulation years) between the two boxes at different seasons, and
the associated standard deviation (std), in the presence of
Lagrangian diffusion. To be compared with table \ref{table:NS}.}
\label{table:NS_diffo}
\end{table}

\begin{figure}[tbp]
\includegraphics[scale=0.5]{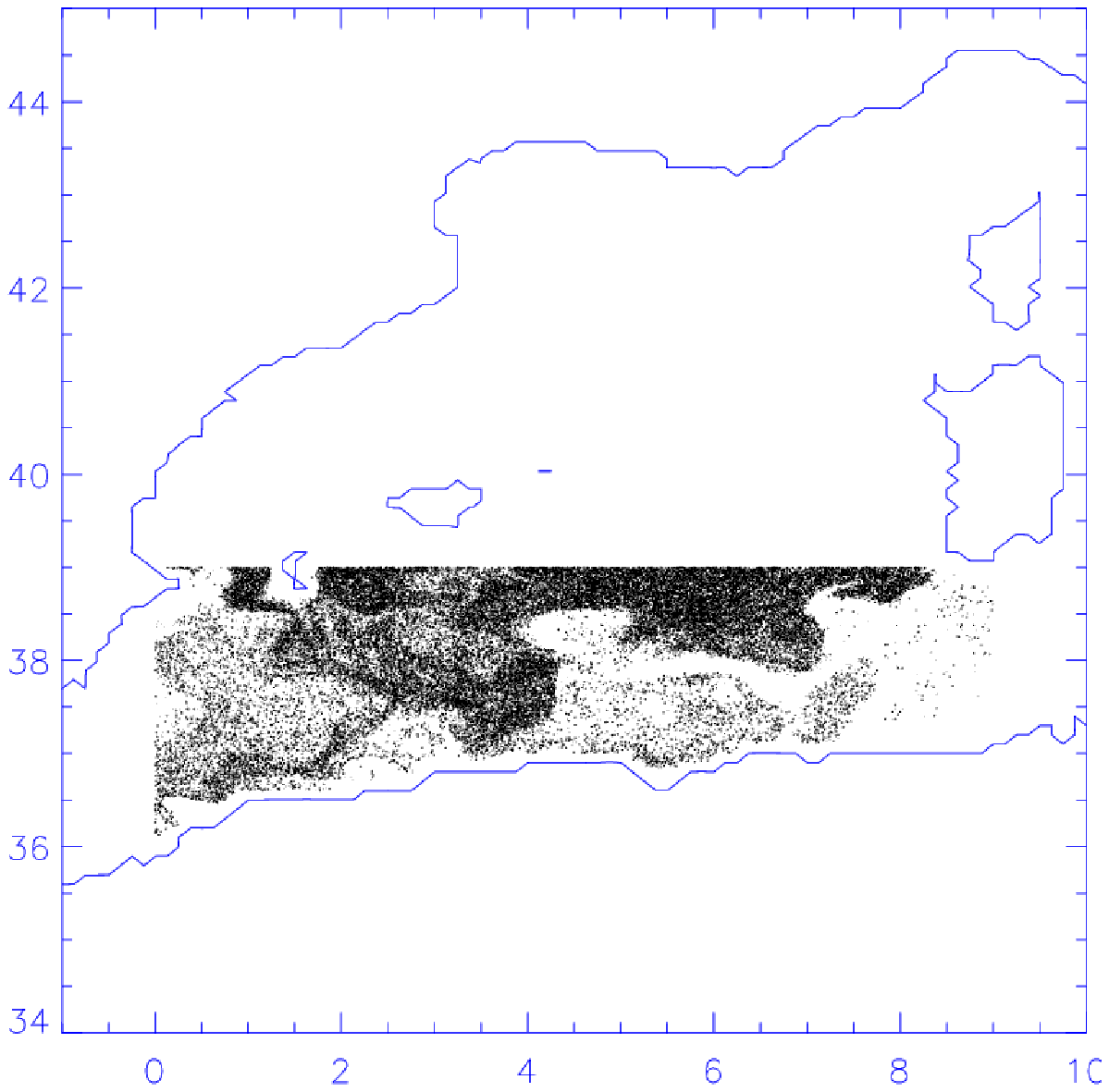}
\includegraphics[scale=0.5]{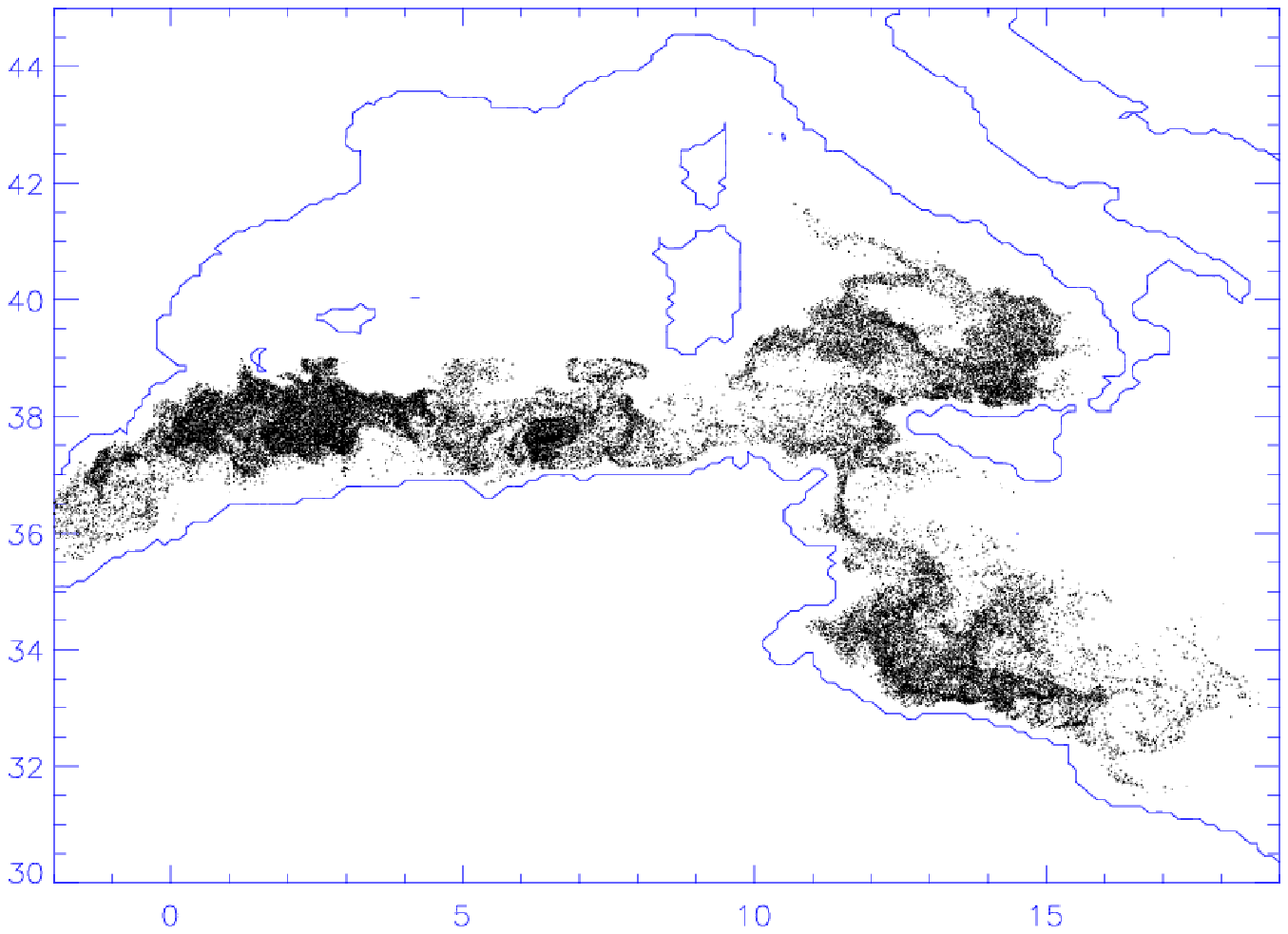}
\caption{\small Geometric structures in the presence of Lagrangian
diffusion. Left: Exchange set towards the north of particles
started in Winter in the southern box. To be compared with the
diffusionless case in the first panel of Fig.~\ref{fig:outS}.
Right: Final positions (after $\tau=90$ days) of particles
launched in Winter in the southern box which did not exchange to
the north. To be compared with the diffusionless case of
Fig.\ref{fig:usmS}.} \label{fig:usmdiffoS}
\end{figure}

In Fig. \ref{fig:usmdiffoS} we show examples of the spatial
Lagrangian structures obtained in the presence of diffusion. We
plot the exchange set towards the northern box starting in Winter
in the southern box (to be compared with the diffusionless first
panel in Fig.~\ref{fig:outS}) and the final positions of particles
started in the southern box and not exchanged with the northern
one (to be compared with the diffusionless Fig. \ref{fig:usmS}).
We see that the main characteristics of the deterministic
evolution are retained. However, the smallest structures,
including the finest filaments, are smoothed out, and this will
occur in all the types of Lagrangian structures visualized. This
does not mean that the sensitive dependence of the fate of
particles to small displacements is lost. At the contrary, since
both exchanged and non-exchanged sets are smoothed-out, they
become still more intimately interleaved (mixed) at the smallest
scales than in the absence of diffusion.

\section{Conclusion}

In this Paper we have analyzed and quantified, from surface velocity fields
provided by a numerical model, several transport processes involving different
water masses in the Western Mediterranean. We have characterized their seasonal
variability by calculating seasonal exchange rates and residence times. The
geometry of the exchange sets and the transport routes has also been
investigated.

Residence times smaller than a season have been found inside a
southern region representing the Algerian basin. Eddies and other
complex structures are involved in the exit process from this
region, being the eastern part of the region the one with fastest
escape.

Exchange of Atlantic waters towards the northern basin occurs at a
higher rate, in the surface layer, than the reverse exchange. In
agreement with \citet{millot}, eddies in the Algerian basin play
an important role in this transport, that is very seasonal and
presents also a component of interannual variability. We expect
the transport of saltier water towards the south to be more
intense in deeper layers, but at surface it involves only the
areas close to the boundary between the regions, and the Catalan
current. Eddy structures are not so prominent for this transport.

Small-scale turbulent process, modelled here as Lagrangian
diffusion, modify the exchange rates in some small amount, but
does not change the qualitative picture. Furthermore, the main
structures and routes of transport remain, although they become
smeared out (implying stronger mixing) at small scales.

The time scales obtained using the present Lagrangian approach
based on a climatological numerical simulation of the
Mediterranean are only a first approximation to the dynamics of
the real Mediterranean. A numerical simulation using more accurate
synoptic atmospheric forcing, including water mass formation, and
a 3D study would give more realistic results. However, we showed
that the actual configuration of the model gives enough complexity
to define filamental structures and routes of transport between
regions in the Mediterranean, as well as to define the complex
transport by mesoscale eddies.

To conclude, we have shown that tools such as the leaking and the
exchange methods are able to identify and characterize dynamical
structures in aperiodic realistic flows, such as our numerically
generated Mediterranean circulation. The methods allow to
visualize the exchange sets and transport routes in a fast and
efficient way. Understanding and good modelling of these
structures is important to improve forecasting in practical
situations: active or passive tracers released on places belonging
to the exchange sets, such as for example an oil spill, will
spread towards the target region.

\section*{Acknowledgements}
 We acknowledge financial support from MCyT (Spain) and FEDER under projects
REN2001-0802-C02-01/MAR (IMAGEN) and BFM2000-1108 (CONOCE), and
from a MCyT-DAAD (Germany) joint program. JS is thankful for
financial support from the International Max Planck Research
School (IMPRS) for Biomimetic Systems in Golm.





\end{document}